\documentclass[12pt]{article}
\usepackage{amsmath} 
\usepackage{amsfonts} 
\begin{document}

\title{On the computation of star products}
\author{A. V. Bratchikov \\
Kuban State Technological University,\\ Krasnodar, 350072,
Russia
}
\date {March 2012}
\maketitle

\begin{abstract}
The problem of finding a generic star product on ${\mathbb R}^n$ is reduced to the computation of a skew-symmetric biderivation. 
\end{abstract}
 


\section{Introduction}

Let $\alpha= 
\alpha^{ij}(x)\partial_i \wedge \partial_j 
$ be a Poisson bivector on ${\mathbb R}^n,$ where $x=(x^1,\ldots,x^n),$ and $x^i,\,i=1,\ldots,n,$ are coordinates.
A star product
is a ${\mathbb R}[[t]]$ linear associative 
product on ${\bf{A}}[[t]],$  defined for $f,g\in {\bf{A}}=C^\infty({\mathbb R}^n)$ by 
\begin{eqnarray} \label {ore}
f\ast g=
\Pi_\ast(f,g), \qquad  \Pi_\ast=\sum_{k\geq 0}t^k \Pi_k,
\end{eqnarray} 
where $\Pi_k$ are bidifferential operators, $\Pi_0(f,g)=fg,$ $\Pi_1=\frac 1 2 \alpha.$  
The operators $\Pi_k$ are determined by the 
associativity equation  
\begin{eqnarray}
\label {asse}
(f\ast g)\ast h=
f\ast (g\ast h)
\end{eqnarray}
for all
$f,g,h\in {\bf{A}}[[t]].$ The star product is the main ingredient of deformation quantization \cite {BFFLS}.

A construction providing a star product associated with any symplectic manifold 
was given in \cite {F}. It allows to compute star products by a simple iterative procedure. 
An formula for the star product on an arbitrary Poisson manifold has been found by Kontsevich \cite {K}. It looks like 
\begin{eqnarray} \label{k}
\ast =F(\alpha),
\end{eqnarray}
where $F$ is described by a series of diagrams. 
However, the problem of finding 
the star product has not been solved, because there is no systematic method to compute weights of the corresponding graphs.

A third order contribution to the generic star product on ${\mathbb R}^n$
was explicitly computed in \cite{PV}. At present, a fourth order 
contribution to the star product is known \cite{KV}.

The problem, which we discuss in the present paper, is related with computation of higher order contributions to star products.
We construct an explicit function $G$ such that    
\begin{eqnarray}
\label {assetu}
\ast =G(\pi(\alpha)),
\end{eqnarray}
for some skew-symmetric biderivation $\pi.$
This formula reduces the problem of finding the $\ast$-product to the computation of a skew-symmetric biderivation. Moreover, it clarifies (somewhat) eq. (\ref{k}).  
To get (\ref{assetu}) we consider a projection of the Maurer-Cartan equation on the image of the Hochschild coboundary operator and find a solution to the projected equation. It defines the function $G.$ The complementary equation determines the part of deformation which depends only on a skew-symmetric biderivation. 

The paper is organized as follows. In section 2, we introduce notations and decompose the Maurer-Cartan equation. In section 3, we construct the function $G(\pi).$

\section {Decomposition of the Maurer-Cartan equation}

Let $C^p({\bf{A}}), p\ge 0, $ be the space of differential $p-$cochains and let ${\delta}$ be the Hochschild coboundary operator 
\begin {eqnarray*}
\delta: C^p({\bf{A}})[[t]]\to C^{p+1}({\bf{A}})[[t]]
\end {eqnarray*}
which is defined by
\begin {multline*} 
\delta \Phi (f_1,f_2,\ldots, f_{p+1})= f_1\Phi (f_2,\ldots, f_{p+1})+\nonumber \\ 
+\sum_{k=1}^p(-1)^k
\Phi (f_1,\ldots, f_{k-1},f_k f_{k+1},f_{k+2},\ldots, f_{p+1})+(-1)^{p+1}\Phi (f_1,\ldots, f_p)f_{p+1}.
\end {multline*}

Equations (\ref {ore}) and (\ref {asse}) together are equivalent to the Maurer-Cartan equation 
\begin {eqnarray}
\delta \Pi=\frac {1}{2} [\Pi,\Pi],
\label {mk}
\end{eqnarray}
where $[\phantom{\Pi},\phantom{\Pi}]$ is the Gerstenhaber bracket. 
For  $\Phi,\Psi\in C^2({\bf{A}})$
\begin {eqnarray*}
[\Phi,\Psi](f,g,h)=\Phi(\Psi(f,g),h)-\Phi(f,\Psi(g,h))+ \Psi(\Phi(f,g),h)- \Psi(f,\Phi(g,h)).
\end{eqnarray*}
In this case \begin {eqnarray*}
[\Phi,\Psi](f,g,h)=[\Psi,\Phi](f,g,h).
\end{eqnarray*}
Let $C^{p}_{nc}({\bf{A}})$ denote the space of differential $p-$cochains vanishing on the constants. We shall seek a solution to eq. (\ref{mk}) satisfying 
\begin {eqnarray*}
\Pi \in C^2_{nc}({\bf{A}}).
\end {eqnarray*} 

Let $\delta^{+}:  C^{p+1}({\bf{A}})[[t]]\to C^{p}({\bf{A}})[[t]]$ be a generalized inverse of $\delta$ 
:
\begin {eqnarray*} \delta\delta^+ \delta =\delta,\qquad
\delta^{+}\delta \delta^{+}=\delta^{{+}}. 
\end {eqnarray*}
The operator $\delta^{+}$ exists \cite{B2}. It is defined by 
\begin {eqnarray*} 
\delta^+=
\lim_{\epsilon \to 0}  \left(\epsilon^2 I+ \delta^T \delta \right)^{-1}\delta^T
=\lim_{\epsilon \to 0} \delta^T \left(\epsilon^2 I+ \delta \delta^T \right)^{-1},
\end {eqnarray*}
where $I$ is the identity map, and $\delta^T: C^{p+1}({\bf A})[[t]]\to C^p({\bf A})[[t]]$ is the transpose of $\delta.$ For \begin {eqnarray*}\Psi=
\Psi_{a_1\ldots a_ {p+1}}(x)X^{a_1}
\otimes \ldots \otimes X^{a_ {p+1}}\in C^{p+1}(A),
\end {eqnarray*} 
where 
\begin {eqnarray*}X^{a}=\frac {1}{{a^1!a^2!\ldots a^n!}}\frac {\partial ^{|a|}} 
{\partial {{(x^1)}^{a^1}} \partial {(x^2)}^{a^2}\ldots \partial {(x^n)}^{a^n}},
\end {eqnarray*}
$a = (a^1, a^2,\ldots, a^n )$ is a multi-index, ${|a |=a^1 + a^2+ \ldots +a^n,}$
we have
\begin {eqnarray*}
\delta^T\Psi=\Psi_{a_1\ldots a_ {p+1}}(x)\delta^T(X^{a_1}
\otimes \ldots \otimes X^{a_ {p+1}}),\phantom{k}
\end {eqnarray*}
\begin {eqnarray*}
 \delta^T (X^{a_1}
\otimes \ldots \otimes X^{a_ {p+1}})=
\sum_{k=1}^p(-1)^{k+1}X^{a_1}\otimes\ldots\otimes\delta^T (X^{a_k}\otimes X^{a_{k+1}})\otimes \ldots \otimes X^{\alpha_{p+1}},
\end {eqnarray*}
\begin {eqnarray*}
\delta^T 
(X^a \otimes  X^b)
=- X^{a+b}+\delta^a_0X^b+X^a \delta^b_0. \phantom{kkkkkkkkkkmmooh}
\end {eqnarray*} 

The restriction of $\delta^+$ to  $C^3_{nc}({\bf{A}})[[t]] $ is given by an infinite series 
\begin {eqnarray*}
\delta^+= \lim_{\epsilon \to 0}
K^{-1}\sum_{m=0}^\infty \left(UK^{-1}\right)^m \delta^T ,
\end {eqnarray*}
where $K = \epsilon^2 I+D,$ 
\begin {eqnarray*}
D(X^a\otimes X^b)=\left(\nu(a)+\nu(b)\right)X^a\otimes X^b, 
\end {eqnarray*}
\begin {eqnarray*}\nu(a)=\prod_{i=1}^n(a^i+1)-2,\end {eqnarray*}
\begin {eqnarray*}
U(X^a\otimes X^b)=\mathop{ {\sum}'}_{s=0}^a X^{a-s}\otimes X^{b+s}
+ \mathop{ {\sum}'}_{s=0}^b X^{a+b-s}\otimes X^s,
\end {eqnarray*}
\begin {eqnarray*}
 \mathop{{\sum}'}_{s=0}^a Y^{s}= \sum_{s=0}^a Y^{s}- Y^{a} -Y^0,\qquad
\sum_{s=0}^a =\sum_{s_1=0}^{a_1}\ldots \sum_{s_p=0}^{a_p}.
\end {eqnarray*}

In accordance with the decomposition 
\begin {eqnarray*}C^3_{nc}({\bf{A}})[[t]]= 
V_1 \oplus 
{
V}_2,\end {eqnarray*}
where 
\begin {eqnarray*} 
V
_1=P C^3_{nc}({\bf{A}})[[t]],
\qquad
{
V}_2= (I-P)C^3_{nc}({\bf{A}})[[t]],\qquad P=\delta\delta^+,
\end{eqnarray*} 
eq. (\ref {mk}) splits as  
\begin {eqnarray}
\delta \Pi=\frac {1}{2} \delta\delta^+[\Pi,\Pi],
\label {mkt}
\end{eqnarray}
\begin{eqnarray} \label {mkr}
(I-\delta\delta^+)[\Pi,\Pi] =0.
\end{eqnarray}
It follows from (\ref{mkt}) that 
\begin {eqnarray}
\Pi= \Upsilon+ \frac {1}{2} \delta^+[\Pi,\Pi],
\label {mks}
\end{eqnarray}
where $\Upsilon \in C^2_{nc}({\bf{A}})[[t]]$
is an arbitrary 
cocycle, $\delta \Upsilon=0,$ subject only to the restriction 
\begin {eqnarray*}
\Upsilon = \frac {t} {2} \alpha +O(t^2).
\end{eqnarray*}            
Equation (\ref {mks}) can be iteratively solved as:
\begin{eqnarray} 
\label{omey}
\Pi = \Upsilon + \frac 1 2 \delta^+[\Upsilon,\Upsilon ]+ 
\ldots.
\end{eqnarray}

\section {
Partial deformation 
}
To obtain an explicit expression for $\Pi$ we introduce the functions 
\begin{eqnarray*}\langle \ldots.  \rangle:
 \left(C^2({\bf {A}})\right)^m 
 \to  {C^2({\bf{A}})},\qquad m=1,2,\ldots,
  \end{eqnarray*}
which recursively 
defined by 
\begin {eqnarray*}\langle \Phi \rangle=\Phi,\qquad \langle \Phi_1,\Phi_2 \rangle=\delta^+[\Phi_1,\Phi_2],
\end{eqnarray*} 
\begin{eqnarray} \label {u}
\langle \Phi_1,\ldots ,\Phi_m \rangle =
\frac 1 2 \sum_{r=1}^{m-1} \sum_{1\leq i_1<\ldots < i_r \leq m } \langle \langle \Phi_{i_1},\ldots,\Phi_{i_r} \rangle,
\langle \Phi_1,\ldots,\widehat{\Phi}_{i_1},\ldots,\widehat{\Phi}_{i_r},\ldots,\Phi_{m} \rangle 
 \rangle
\end{eqnarray} 
if $m=3,4,\ldots,$ where $\widehat{\Phi}$ means that ${\Phi}$ is omitted.
Equation (\ref {mks}) can be written as 
\begin{eqnarray} 
\label{mks1}
\Pi = \Upsilon + \frac 1 2 \langle \Pi,\Pi \rangle.  
\end{eqnarray}
Using induction on $m$ one easily verifies that $ \langle \Phi_1,\ldots ,\Phi_m \rangle$ is an $m-$linear symmetric function.

For $m\geq 2,1\leq i,j\leq m,$ let  
\begin {eqnarray*}P^m_{ij}:  (C^2({\bf{A}}))^m\to (C^2({\bf{A}}))^{m-1}
\end{eqnarray*}   
be defined by
\begin{eqnarray*} 
P^m_{ij}( \Phi_1,\ldots ,\Phi_m ) = 
( \langle \Phi_i,{\Phi}_{j}\rangle ,\Phi_1 ,\ldots,\widehat{\Phi}_{i},\ldots,\widehat{\Phi}_{j},\ldots,\Phi_{m} ). 
\end{eqnarray*} 
If $\Phi \in C^2({\bf{A}})$ is given by  
\begin{eqnarray} 
 \Phi = P^2_{12}P^3_{
i_{m-2}j_{m-2}
}\ldots P^{m-1}_
{i_2j_2}
P^m_{i_1j_1}
(\Phi_1,\ldots ,\Phi_m )
\label {ku}
\end{eqnarray} 
for some $ (i_1j_1),\ldots,(i_{m-2}j_{m-2}),$ we say that $\Phi$ is a descendant of $(\Phi_1,\ldots ,\Phi_m ).
$
A descendant of $\Phi\in C^2({\bf {A}})$ is defined as $\Phi.$
One can show that $\langle \Phi_1,\ldots,\Phi_m \rangle$ equals the sum of all the descendants of  $(\Phi_1,\ldots ,\Phi_m )$ \cite {B3}.
\vspace{1mm}
For example, \begin{eqnarray*} \langle \Phi_1,\Phi_2,\Phi_{3}\rangle=
\langle \langle \Phi_{1},\Phi_{2}\rangle, \Phi_3 \rangle+\langle \langle \Phi_{1},\Phi_{3}\rangle, \Phi_2 \rangle+\langle  \langle \Phi_2, \Phi_3 \rangle,\Phi_{1} \rangle. 
\end{eqnarray*} 
The function 
$\langle \ldots.  \rangle:
 \left(C^2({\bf {A}})\right)^m 
 \to  {C^2({\bf{A}})}$
can be uniquely extended to a function $\langle \ldots.  \rangle:\left(C^2({\bf {A}})[[t]]\right)^m\to  {C^2({\bf{A}})[[t]]}$ by ${\mathbb R}[[t]]$-linearity. 

Equation (\ref {omey}) can be written as \cite{B3}
\begin{eqnarray} \label {orro}
\Pi= \langle e^{\Upsilon} \rangle,
\end{eqnarray}
where
\begin {eqnarray*} \langle e^{\Upsilon} \rangle =\sum_{m\geq 0} \frac {1} {m!}\langle \Upsilon^m \rangle 
, \qquad \langle \Upsilon^0 \rangle=0.
\end{eqnarray*} 
Substituting (\ref{orro}) into (\ref{mkr}), we get the equation determining $ \Upsilon:$
\begin{eqnarray} \label{lll} 
(I-\delta\delta^+)[\langle e^{\Upsilon} \rangle,\langle e^{\Upsilon} \rangle] =0.
\end{eqnarray}
The cocycle $\Upsilon$ can be written as $\Upsilon= \pi+ \delta \lambda,$ where  $\pi=\sum_{k\geq 1}t^k \pi_k$ is a skew-symmetric biderivation, and $\lambda$ is an $1-$cochain \cite {GR}. 

Let us assume by induction that $\lambda=\sum_{k\geq n} t^k \lambda_k.$ Then one can write
\begin{eqnarray} \label {orrox}
\Pi_\ast = \Pi_0+\langle e^{\Upsilon} \rangle= \sum_{k=0}^{n-1}t^k \Pi_k + t^n(\pi_n+ \delta \lambda_n + \tilde \Pi_n)+ O(t^{n+1})
\end{eqnarray}
where $\tilde \Pi_n$ depends only on $\pi_k$ with $k<n.$ 
Define a map $S: {\bf A}[[t]]\to {\bf A}[[t]]$ by
\begin {eqnarray*}S(f)= f - t^n\lambda_n(f)
\end{eqnarray*} 
with inverse
\begin {eqnarray*}S^{-1}(f)= f + \sum_{k\geq 1} t^{kn}\lambda_n^k(f)
\end{eqnarray*} 
for any $f\in {\bf A}.$
Then 
\begin{eqnarray} \label {oms}
\Pi'_\ast(f,g)= S^{-1}(\Pi_\ast(S(f),S(g)))
\end{eqnarray} 
is a new equivalent star product
\begin {eqnarray*}\Pi'_\ast=\Pi_0+ \langle e^{\Upsilon'} \rangle, \qquad \delta\Upsilon'=0.
\end{eqnarray*}  
It follows from (\ref{orrox}) and (\ref{oms}) that 
\begin {eqnarray*}{\Upsilon'}= \sum_{k=1}^n t^{k}\pi_k+O(t^{n+1}).
\end {eqnarray*}
It then follows from the vanishing of $\delta\lambda_n$ that the coboundary  $\delta\lambda$ can be removed from (\ref{orro}) by a similarity transformation. 

Finally, we get
\begin{eqnarray} \label {orrod}
\Pi= \langle e^{\pi} \rangle, \qquad \pi = \frac t 2 \alpha + \sum_{k\geq 2}^\infty t^k\pi_{k},
\end{eqnarray}
where $\pi_k=\pi_k^{ij}\partial_i \wedge \partial_j.$ Equation (\ref{lll}) takes the form  
\begin{eqnarray*} 
(I-\delta\delta^+)[\langle e^{\pi} \rangle,\langle e^{\pi} \rangle] =0.
\end{eqnarray*}
The cocycle $\pi$ depends on $\alpha^{ij}$ and the derivatives $ \partial_{k_1} \ldots  \partial_{k_m}\alpha^{ij},$ ${m=1,2\ldots.}$ 
In eq. (\ref{assetu}) 
\begin{eqnarray*} G(\pi)=\Pi_0+\langle e^{\pi} \rangle.
\end{eqnarray*} 

\section {Conclutions}
In this paper we have studied a construction of the generic star product on ${\mathbb R}^n.$ Our approach is based on a decomposition of the Maurer-Cartan equation. In accordance with this decomposition the star product is represented as a composition of two functions.
We show that one of these functions is a skew-symmetric biderivation and give an explicit expression for the other. It is an important problem to find a solution to the equation determining the skew-symmetric biderivation. 
It would also be interesting to reproduce decomposition (\ref{assetu}) in the framework of the path integral approach \cite{CF}.


\begin{thebibliography}{}
\bibitem{BFFLS} F. Bayen, M. Flato, C. Fronsdal, A. Lichnerovich and D. Sternheimer, 
Deformation theory and quantization,
{\it Ann. Phys.} {\bf 111}  (1978), 61-110.
\bibitem{F} B. Fedosov, A simple geometrical construction of deformation quantization, 
{\it J. Diff. Geom.} {\bf 40} (1994), 213-238.
\bibitem{K} M. Kontsevich, Deformation quantization of Poisson manifolds,
{\it Lett. Math. Phys.} {\bf 66}  (2003), 157-216.
\bibitem{PV} M. Penkava and P. Vanhaecke, Deformation quantization of polynomial Poisson algebras,
{\it J. Algebra } {\bf 227}  (2000), 365-393.
\bibitem{KV} V. G. Kupriyanov and D. V. Vassilevich, Star products made (somewhat) easier,
{\it Eur. Phys. J. C } {\bf 58} (2008), 627-637. 
\bibitem{B2} A. V. Bratchikov, Generalized inversion of 
the Hochschild  coboundary operator and deformation quantization,
{\it Int. J. Geom. Meth. Mod. Phys.} {\bf 8} (2011), 99-106.
\bibitem{B3} A. V. Bratchikov, Explicit construction of the classical BRST charge for nonlinear algebras, {\it Central Eur. J. Phys.} {\bf 10} (2012), 61-65.
\bibitem{GR} S. Gutt and J. Rawnsley, Equivalence of star products on a symplectic manifold, 
{\it J. Geom. Phys.} {\bf 29} (1999), 347-392.
\bibitem{CF} A. C. Cattaneo and G. Felder, A path integral approach to the Kontsevich
quantization formula, {\it Commun. Math. Phys.} {\bf 212} (2000), 591-611. 
\end{thebibliography}
\end{document}